\documentclass[journal]{IEEEtran}
\def\@IEEEclspkgerror{\ClassError{IEEEtran}}
\usepackage{enumitem}
\usepackage{cite}
\usepackage{graphicx}
\usepackage{algorithm}
\usepackage{amsmath}
\usepackage{amssymb}
\usepackage{multirow}

\usepackage{algpseudocode}
\usepackage{amsthm}

\usepackage{soul}
\usepackage[dvipsnames]{xcolor}
\usepackage{hyperref}
\usepackage[dvipsnames]{xcolor}

\usepackage{array}

\makeatletter
\newcounter{parenttheorem}

\makeatother

\def\BibTeX{{\rm B\kern-.05em{\sc i\kern-.025em b}\kern-.08em
    T\kern-.1667em\lower.7ex\hbox{E}\kern-.125emX}}
\usepackage{lineno}

\begin{document}

\title{Surface Recognition for e-Scooter \\ Using Smartphone IMU Sensor}
\author{Areej Eweida, Nimord Segol, Maxim Freydin, \IEEEmembership{Member, IEEE}, Niv Sfaradi, and Barak Or, \IEEEmembership{Member, IEEE}
    \thanks{Submitted: Jan 2023.}
\thanks{All authors are with ALMA Technologies Ltd, Haifa, 340000, Israel (e-mail: barak@almatechnologies.com).}}

\markboth{Surface Recognition for e-Scooter Using Smartphone IMU Sensor PREPRINT}%
{TBD}
\maketitle

\begin{abstract}

In recent years, as the use of micromobility gained popularity, technological  challenges  connected to e-scooters became increasingly important. This paper focuses on
road surface recognition, an important task in this area. A reliable and accurate method for road surface recognition can help improve the safety and stability of the vehicle. Here a data-driven method is proposed to recognize if an e-scooter is on a road or a sidewalk.  The proposed method uses only the widely available inertial measurement unit (IMU) sensors on a smartphone device.  deep neural networks (DNNs) are used to infer whether an e-scooter  is driving on a road or on a  sidewalk by solving a binary classification problem. A data set is collected and several different deep models as well as classical machine learning approaches for the binary classification problem are applied and compared. Experiment results on a route containing the two surfaces are presented demonstrating the DNNs ability to distinguish between them.  
\end{abstract}


\section{Introduction}\label{sec:introduction}

\IEEEPARstart{R}{oad} surface recognition is the process of identifying and classifying different types of road surfaces that a vehicle may encounter during operation \cite{du2020abnormal,bystrov2015analysis,aki2016road,cafiso2022urban}. This work focuses on road surface recognition for e-scooters. A mode of transportation that has become increasingly popular in urban areas in recent years. In particular, this work presents a method for distinguishing between paved asphalt roads and sidewalks for e-scooters using only a mobile device. 

Accurate road surface recognition is important for several reasons. First, different road surfaces can have different impacts on the stability and handling of an e-scooter \cite{leoni2022assessing}. For example, a smooth, paved road may provide a stable and comfortable ride, while a textured, rough sidewalk may be more difficult to navigate. By identifying the road surface, e-scooters can adjust their performance accordingly, ensuring a safe and enjoyable ride for the user. 
Additionally, road surface recognition can be used to improve the efficiency of e-scooters. Different road surfaces can have different levels of resistance, which can impact the amount of energy required to travel a given distance. By accurately identifying the road surface, e-scooters can optimize their power output and energy consumption, extending the range of the vehicle and improving the overall user experience. Finally, driving an e-scooter on a sidewalk is illegal in many places. An accurate method that determines if a rider is on a sidewalk or on a road can provide fleet managers and micromobility companies with a valuable tool.

In recent years, several approaches  were developed to address the problem of road surface recognition for different transportation modes. 
In\cite{Springer2020} the authors present a road surface recognition system for bicycles using several sensors attached to the bicycle including a camera an inertial measurement unit (IMU) and a speedometer. 
In  \cite{Chen_2022}, the authors described a method to estimate the roughness of a surface on which a  car is traversing using LiDAR as well as a global positioning system (GPS) receiver and IMU. 
In \cite{road_surface_rec2019} the authors describe a method of surface recognition for pedestrians using an IMU and a microphone attached to the users' shoes.
A method to estimate the condition of the road surface using lasers mounted on a vehicle was proposed in \cite{laser_scanners2019}.
In \cite{road_surface_dl2020} a deep classifier that estimates the road surface was trained using data collected by cameras. In \cite{du2020abnormal} an algorithm is presented that extracts features from an IMU sensor on a car and uses K nearest neighbors algorithm to detect abnormal road surface conditions. 

Other related works where machine learning  algorithms, and in particular deep neural networks, were applied to IMU sensor data include \cite{liu2021vehicle} where DNN-based multi-models together with an EKF were proposed to deal with the global navigation satellite system (GNSS) outage.  In \cite{or2022hybrid} it was demonstrated that integration of DNN into classical signal
processing methods can boost navigation performance. In \cite{freydin2022mountnet} the authors demonstrate how a DNN model can estimate a mounting yaw angle of a smartphone strapped to a vehicle. Another related work is \cite{yan2018ridi} where DNNs were used to learn linear velocities from IMU sensors in cars.
To the best of the author's knowledge, this work is the first to present a road surface recognition algorithm, including a DNN, for e-scooters.

In this paper, a novel approach to road surface recognition for e-scooters using IMU sensors on smartphones is developed. The method is data-driven and relies on a DNN binary classifier. IMU sensors are widely available in smartphones. By mounting a smartphone on an e-scooter and collecting data from the IMU sensors, and applying a DNN-based algorithm, it is possible to infer the type of road surface the e-scooter is traveling on. The DNN is able to detect the signature associated with each type of road surface and discern between roads and sidewalks.
The proposed approach has several advantages: It is cost-effective, as it utilizes sensors that are already present in many smartphones; It is also lightweight and compact, making it well-suited for use on e-scooters; Additionally, smartphones are increasingly being used as onboard computers for e-scooters, providing a convenient platform for integrating the suggested road surface recognition system.

The rest of the paper is organized as follows: In Section \ref{sec:Dateset Creation} the data collection and dataset generation processes are presented. In Section \ref{sec:model} the proposed ML algorithms are detailed. In Section \ref{sec:results} the experimental results are described. In particular, the results of several algorithms on a validation set are presented. An experiment on a route containing both types of surfaces is also presented. Finally, in Section \ref{sec:conclusions} conclusions are given. 

\section{Data set creation}
\label{sec:Dateset Creation}
Any data-driven method requires an offline collection of data and model training. This section contains a description of the data collection and processing.  A schematic view of the data process can be seen in Fig. \eqref{fig:data_gen_schem}.

\subsection{Raw data collection}
\begin{figure}[ht]
\centering
{\includegraphics[width=0.47\textwidth]{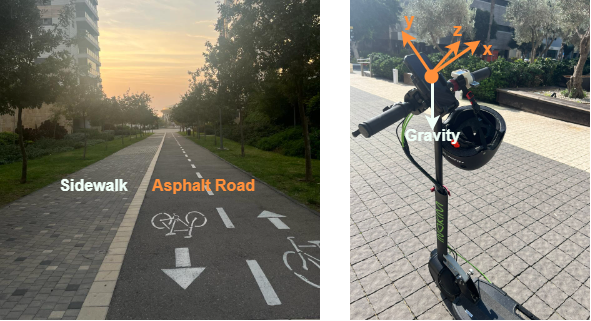}}
\caption{Left: an example of a street with both sidewalk and an asphalt road. Right: a smartphone device mounted on a scooter. Note the device frame of reference.}
\label{fig:data_collection_road}
\label{fig:Smartphone}
\end{figure}

IMU acceleration and gyroscope data was collected with an iPhone 13 strapped to an e-scooter's handle. Data was recorded at the maximum rate of 100 [Hz]. A total of 100 minutes were collected for this task in 12 distinct driving sessions conducted by the same rider. Different driving sessions had distinct routes to increase diversity in the data set. See Fig. \eqref{fig:data_collection_road} for an image of the route and the collecting device (smartphone).
 Of the routes, 7 were entirely on a sidewalk, and the rest were on the road. To create a validation set, 2 of the drives were set aside. All in all, 10 drives totaling 80 minutes were used for training and the two drives, one on a sidewalk and one on the road, for validation. 

\subsection{Pre-processing}
\label{sec:processing}
The collected data consists of acceleration $[m/s^{2}]$ and angular velocity $[rad/s]$ in three axes. Before passing the acceleration and gyroscope data to the classifier, the following pre-processing tasks are applied:
\begin{enumerate}
\item Apply a low-pass filter with a cut-off frequency of 
10 [Hz].
\item Down-sample the data to 20 [Hz].
\item Rotate (only roll and pitch) the measured acceleration and angular velocity vectors to the navigation frame where the first component points to the e-scooter forward direction and the third component is parallel to the gravity vector.
\item Subtract the gravity component from the third acceleration component.
\end{enumerate}

\begin{figure}[ht]
\centering
{\includegraphics[width=0.4\textwidth]{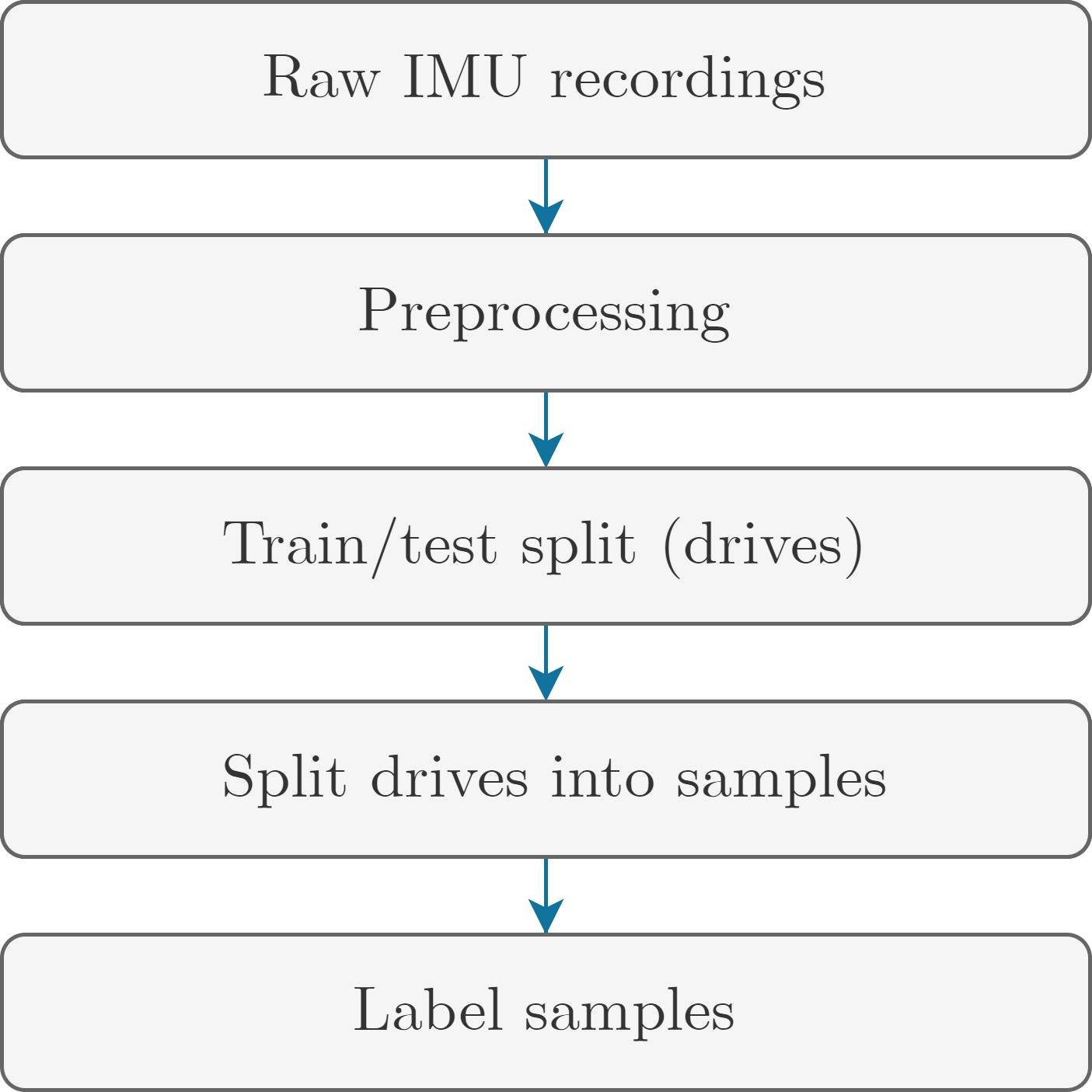}}
\caption{Schematic description of the dataset generation process.}
\label{fig:data_creation}
\label{fig:data_gen_schem}
\end{figure}

These  steps are applied both offline when training the model and online as a part of the classification algorithm. The goal of the described processing is to simplify the signal and avoid unnecessary noise and other information that can affect the performance of the classifier.
It is noted that the above processing procedure has appeared in various prior works e.g. \cite{farrell2008aided, freydin2022learning, freydin2022mountnet}. It is provided in Figure \ref{fig:data_creation} for this work to be self-contained. 

\subsection{Data generation}
\label{sec:data_generation}

To create the dataset the recorded drives were partitioned to windows of $S$ seconds. In this work the value of $S$ was considered a hyper-parameter in the set $\{1, 2, 3\}$.
The overall data set consists of pairs $X\in \mathrm{R}^{20\cdot S\times 6}$  of IMU  measurements of $S[s]$ sampled in 
 20 [Hz] and $Y\in\{0,1\}$ with $1$ indicating a sidewalk and $0$ indicating a paved road. 

\
\section{Classifiers}
\label{sec:model}
Using the dataset collected as described above we train a total of 5 ML models. Each of the models outputs one or zero signifying sidewalk or road.
\subsection{Models} 
\label{sec:models}
Five ML models where tested. The most common DNN-based models for this type of task are convolutional neural network (CNN) \cite{o2015introduction} and long short-term memory (LSTM) \cite{hochreiter1997long}. Each one of them were used as a basis for a DNN architecture and both of them were combined in a fused architecture. In addition, two classical ML models were tested: support vector machine (SVM) \cite{cortes1995support} and random forest \cite{breiman2001random}. A description of each model is provided: 
\begin{enumerate}
    \item  {\bf CNN} - A convolutional neural network with two convolution layers with 8 channels and 3 kernels followed by dropout and mean pooling with a rectified linear unit (ReLU). One convolution layer followed by dropout and max pooling with ReLU. Two linear layers with ReLU with 20 and 1 channels, respectively.
    \item {\bf LSTM} - A simplified LSTM layer with hidden size of 100 followed by a dense layer.
    \item {\bf LSTM-CNN} - An LSTM layer with hidden size 100 followed by two convolution layers with 8 channels and 3 kernels each, followed by dropout and mean pooling with ReLU. The model ends with a single linear layer, as presented in Fig.\ref{fig:CNN}.
    
    \item {\bf SVM} -  Each sample $X\in\mathbb{R}^{20\cdot S\times 6}$ was flattened to a vector $\Tilde{X}\in \mathbb{R}^{120\cdot S}$ and a classifier was trained on it.
    \item {\bf Random Forest} - Each sample $X\in\mathbb{R}^{20\cdot S\times 6}$ was flattened to a vector $\Tilde{X}\in \mathbb{R}^{120\cdot S}$ and an ensemble random forest classifier was trained. The maximum depth allowed per tree was set to 100.
\end{enumerate}

\subsection{DNNs loss function and training hyperparameters}
The loss function and training criteria for the DNN-based models are described in this section.
\subsubsection{Loss function for DNN-based models}
All of the DNNs considered in this work were trained using the binary cross entropy loss, the most common loss function used for binary classification problems. It is calculated using the following formula:
\begin{equation}
\ell(p_i, Y_i)  = -\left[{ Y_i \cdot \log{(p_i)} + (1-Y_i) \cdot \log{(1-p_i)} }\right],
\end{equation}
where the output of the DNN is $p_i= p (X_i, {\bf{\Theta})}\in [0,1]$, which depends on the sample $X_i$ and the weights $\bf{\Theta}$ and $Y_i\in \{0,1\}$ is the true label of example $i$.
The entire loss is 
\begin{equation}
L  =  \frac1N\sum_{i=1}^{N}\ell(p_i, Y_i),
\end{equation}
where $N$ is the number of samples in the train set.

\subsubsection{Training DNN-based models}
The DNN-based models were all trained for 30 epochs using the ADAM optimizer \cite{kingma2014adam} with a learning rate of $0.001$, $\beta$ coefficients
of $\beta_1=0.9$, $\beta_2=0.999$, and batch size $6$. We trained the model using a Tesla K80 GPU. We trained for
approximately 3.5 minutes on average in each combination of architecture and window size, 8 seconds per epoch.

\section{Results}
This section contains the results of each of the considered classifiers on the validation set. The models were evaluated for windows of $S=\{1, 2, 3\}$ seconds. The model which obtained the best results on the validation set was then tested on a separate drive containing both road and sidewalk.
\label{sec:results}
\subsection{Evaluation Metrics}
Three evaluation metrics for the performance of the models were considered. 
\begin{enumerate}
\item {\bf Accuracy}: accuracy is the proportion of true results among the total number of cases examined.
\item {\bf F1 Score}: the F1 score is a number between 0 and 1 and represents the harmonic mean of precision and recall.
\item {\bf False Positive Rate (FPR)}: FPR is the proportion of samples classified as 1 when the true label was 0.
\end{enumerate}
Each of these metrics was computed for the validation set for all the ML models.

\begin{table}[ht]
\centering
\begin{tabular}{| m{1.8cm} | m{1cm}| m{1cm} | m{1cm} |  m{1cm} |} 
\hline
\textbf{Model} & \textbf{Window Size} [s] & \textbf{Accuracy} & \textbf{F1}  & \textbf{FPR} \\ 
\hline
\multirow{3}{*}{CNN}  
            &     1     &     0.78      & 0.72    &  0.22    \\
\cline{2-5}
            &     2     &     0.83     & 0.83    & 0.15    \\
\cline{2-5}
            &     3     &     0.78     & 0.78    & 0.16    \\
\hline
\multirow{3}{*}{LSTM}  
            &     1     &     0.86     & 0.88    & 0.15    \\
\cline{2-5}
            &     2     &     0.85      & 0.86    & 0.17    \\
\cline{2-5}
            &     3    &     0.52     & 0.47    & 0.23    \\
\hline
\multirow{3}{*}{LSTM-CNN}  
            &     1    &     0.88     & 0.88    & 0.14    \\
\cline{2-5}
            &     2    &     0.90      & 0.90    & 0.13    \\
\cline{2-5}
            &     3   &     \bf{0.93}     &  \bf{0.93}    & \bf{0.13}    \\
\hline

\multirow{3}{*}{SVM}  
            &     1    &     0.73     & 0.70    & 0.16    \\
\cline{2-5}
            &     2    &     0.72      & 0.68    & 0.16    \\
\cline{2-5}
            &     3    &     0.71      & 0.66    & 0.18    \\
\hline
\multirow{3}{*}{Random Forest}  
            &     1   &     0.79      & 0.79    & 0.23    \\
\cline{2-5}
            &     2  &     0.78      & 0.79    & 0.22    \\
\cline{2-5}
            &     3    &     0.80      & 0.82    & 0.20    \\
\hline

\end{tabular}
\vspace{0.2cm}
\caption{Performance metrics for each model on the validation set for different window sizes.}
\label{Tab:results}
\end{table}

\subsection{Best performing model}
\begin{figure}[!ht]
\centering
{\includegraphics[width=0.5\textwidth]{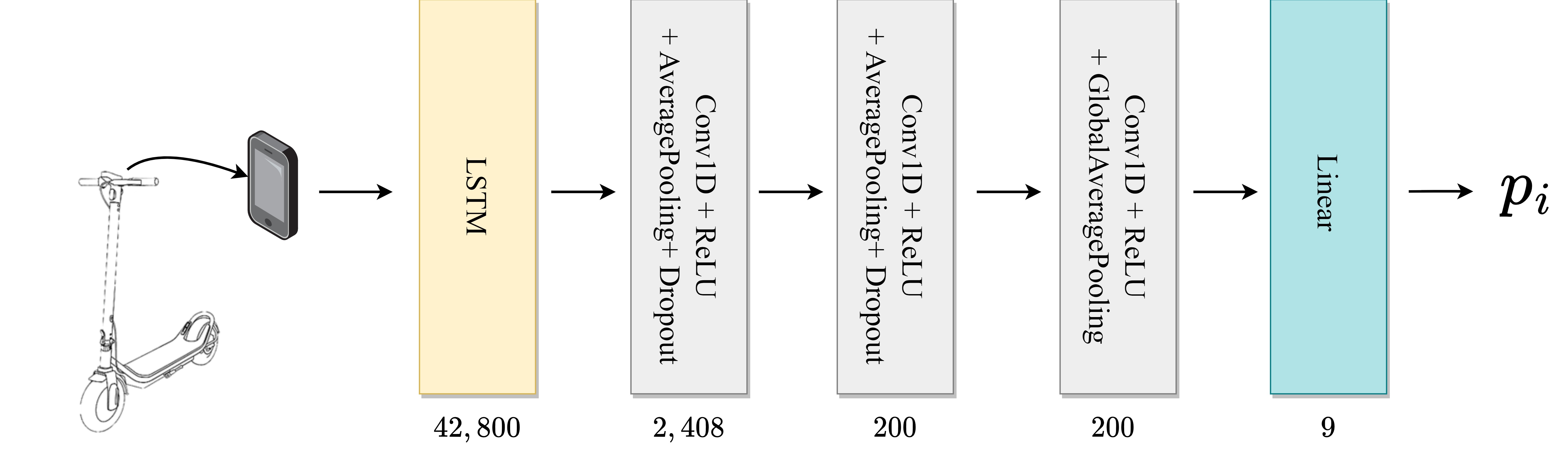}}
\caption{DNN architecture: LSTM, two convolutional layers are followed by one dense layer with a connecting average-pooling layer. Finally, a linear layer eventually outputs the probability of getting each one of the classes.}
\label{fig:CNN}
\end{figure}

The optimal solution for the classification problem on the validation set is the LSTM-CNN model. The benefit of this model is its both spatially (IMU signals information) and temporally (time) properties. The architecture layers are presented in Fig. \ref{fig:CNN}. The best results on the validation set were obtained for a window of $3[s]$. Table \ref{Tab:results} presents performance metrics for each of the ML models considered in this work. Note that the model outperforms all other models where it achieves maximal accuracy of $0.93$, maximal F1 of $0.93$, and minimal FPR of $0.13$. 

\subsection{The performance of classical models}
The two classical models considered were the SVM and Random Forest. In contrast to the LSTM and CNN-based models, the classical approaches do not treat the input signal as time series but as a flattened vector of features. Table \ref{Tab:results} shows a 
surprising result that both classical models performed well and closely approached the metrics of the DNNs. This may suggest that the data is separable as is even without significant effort in handcrafting features. However, it is important to note that the IMU data was filtered, downsampled, rotated, and the gravity component was subtracted.

\subsection{Field experiment}
\begin{figure}[ht]
\centering
{\includegraphics[width=0.48\textwidth]{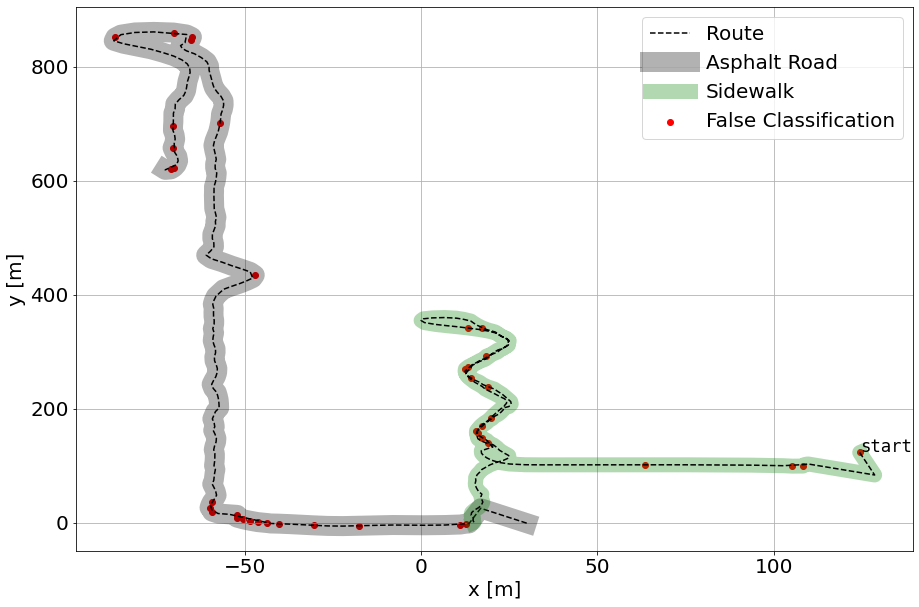}}
\caption{Results of the selected classifier on a route containing both sidewalk and road.}
\label{fig:route}
\label{fig:expermient}
\end{figure}

An experiment was conducted on a single drive containing both road and sidewalk segments. The route is shown in Fig. \eqref{fig:expermient} and it was not used in the training or validation sets. The light green segment is a sidewalk, the light gray is an asphalt road, and the red points show false classification. The route begins on a sidewalk and at a certain point, switches to an asphalt road. 

The trained LSTM-CNN model with a $3[s]$ window was chosen for this task as it performed best on the validation set. A recording of raw IMU measurement was passed through the pre-processing described in Section \ref{sec:processing} and fed to the classifier. The model was activated every $5[s]$. The classifier has successfully classified $90\%$ of the route. In addition, a strong wind was present during the recording of this experiment, which was not the case for the training data. The wind has led to elevated levels of noise and vibration which were not present in the training or validation datasets. However, the results of this experiment agree well with the performance observed on the validation set.



\section{Conclusions}
\label{sec:conclusions}
A data-driven approach to the road surface recognition problem for e-scooters was suggested. For e-scooters it is specifically important to distinguish between roads and sidewalks. A binary classifier was trained to perform this task. Several models both based on deep learning and classical approaches were tested. Experimental results showed that a DNN-based model produces good results. Future research directions include developing a multi-class classifier that distinguishes between several types of surfaces and developing a real-time system that incorporates data from several sensors using a mobile device to deduce the road surface.

\bibliographystyle{IEEEtran}
\bibliography{IEEEfull}

\begin{thebibliography}{10}
\providecommand{\url}[1]{#1}
\csname url@samestyle\endcsname
\providecommand{\newblock}{\relax}
\providecommand{\bibinfo}[2]{#2}
\providecommand{\BIBentrySTDinterwordspacing}{\spaceskip=0pt\relax}
\providecommand{\BIBentryALTinterwordstretchfactor}{4}
\providecommand{\BIBentryALTinterwordspacing}{\spaceskip=\fontdimen2\font plus
\BIBentryALTinterwordstretchfactor\fontdimen3\font minus
  \fontdimen4\font\relax}
\providecommand{\BIBforeignlanguage}[2]{{%
\expandafter\ifx\csname l@#1\endcsname\relax
\typeout{** WARNING: IEEEtran.bst: No hyphenation pattern has been}%
\typeout{** loaded for the language `#1'. Using the pattern for}%
\typeout{** the default language instead.}%
\else
\language=\csname l@#1\endcsname
\fi
#2}}
\providecommand{\BIBdecl}{\relax}
\BIBdecl

\bibitem{du2020abnormal}
R.~Du, G.~Qiu, K.~Gao, L.~Hu, and L.~Liu, ``Abnormal road surface recognition
  based on smartphone acceleration sensor,'' \emph{Sensors}, vol.~20, no.~2, p.
  451, 2020.

\bibitem{bystrov2015analysis}
A.~Bystrov, M.~Abbas, E.~Hoare, T.-Y. Tran, N.~Clarke, M.~Gashinova, and
  M.~Cherniakov, ``Analysis of classification algorithms applied to road
  surface recognition,'' in \emph{2015 IEEE Radar Conference (RadarCon)}.\hskip
  1em plus 0.5em minus 0.4em\relax IEEE, 2015, pp. 0907--0911.

\bibitem{aki2016road}
M.~Aki, T.~Rojanaarpa, K.~Nakano, Y.~Suda, N.~Takasuka, T.~Isogai, and
  T.~Kawai, ``Road surface recognition using laser radar for automatic
  platooning,'' \emph{IEEE Transactions on Intelligent Transportation Systems},
  vol.~17, no.~10, pp. 2800--2810, 2016.

\bibitem{cafiso2022urban}
S.~Cafiso, A.~Di~Graziano, V.~Marchetta, and G.~Pappalardo, ``Urban road
  pavements monitoring and assessment using bike and e-scooter as probe
  vehicles,'' \emph{Case Studies in Construction Materials}, vol.~16, p.
  e00889, 2022.

\bibitem{leoni2022assessing}
J.~Leoni, M.~Tanelli, S.~C. Strada, and S.~M. Savaresi, ``Assessing e-scooters
  safety and drivability: a quantitative analysis,'' \emph{IFAC-PapersOnLine},
  vol.~55, no.~24, pp. 260--265, 2022.

\bibitem{Springer2020}
M.~Springer and C.~Ament, ``A mobile and modular low-cost sensor system for
  road surface recognition using a bicycle,'' in \emph{2020 IEEE International
  Conference on Multisensor Fusion and Integration for Intelligent Systems
  (MFI)}, 2020, pp. 360--366.

\bibitem{Chen_2022}
K.~Chen, ``Road roughness recognition based on lidar,'' \emph{Journal of
  Physics: Conference Series}, vol. 2278, no.~1, p. 012008, may 2022.

\bibitem{road_surface_rec2019}
H.~Mitake, H.~Watanabe, and M.~Sugimoto, ``Footsteps and inertial data-based
  road surface condition recognition method,'' in \emph{Proceedings of the 18th
  International Conference on Mobile and Ubiquitous Multimedia}, ser. MUM
  '19.\hskip 1em plus 0.5em minus 0.4em\relax New York, NY, USA: Association
  for Computing Machinery, 2019.

\bibitem{laser_scanners2019}
K.~Urano, K.~Hiroi, S.~Kato, N.~Komagata, and N.~Kawaguchi, ``Road surface
  condition inspection using a laser scanner mounted on an autonomous driving
  car,'' in \emph{2019 IEEE International Conference on Pervasive Computing and
  Communications Workshops (PerCom Workshops)}, 2019, pp. 826--831.

\bibitem{road_surface_dl2020}
E.~Šabanovič, V.~Žuraulis, O.~Prentkovskis, and V.~Skrickij,
  ``Identification of road-surface type using deep neural networks for friction
  coefficient estimation,'' \emph{Sensors}, vol.~20, no.~3, 2020.

\bibitem{liu2021vehicle}
J.~Liu and G.~Guo, ``Vehicle localization during gps outages with extended
  kalman filter and deep learning,'' \emph{IEEE Transactions on Instrumentation
  and Measurement}, vol.~70, pp. 1--10, 2021.

\bibitem{or2022hybrid}
B.~{O}r and I.~Klein, ``A {H}ybrid {M}odel and {L}earning-{B}ased {A}daptive
  {N}avigation {F}ilter,'' \emph{IEEE Transactions on Instrumentation and
  Measurement}, pp. 1--1, 2022.

\bibitem{freydin2022mountnet}
M.~Freydin, N.~Sfaradi, N.~Segol, A.~Eweida, and B.~Or, ``Mountnet: Learning an
  inertial sensor mounting angle with deep neural networks,'' \emph{arXiv
  preprint arXiv:2212.11120}, 2022.

\bibitem{yan2018ridi}
H.~Yan, Q.~Shan, and Y.~Furukawa, ``{RIDI}: Robust {IMU} double integration,''
  in \emph{Proceedings of the European Conference on Computer Vision (ECCV)},
  2018, pp. 621--636.

\bibitem{farrell2008aided}
J.~Farrell, \emph{Aided navigation: GPS with high rate sensors}.\hskip 1em plus
  0.5em minus 0.4em\relax McGraw-Hill, Inc., 2008.

\bibitem{freydin2022learning}
M.~Freydin and B.~Or, ``Learning car speed using inertial sensors for dead
  reckoning navigation,'' \emph{IEEE Sensors Letters}, vol.~6, no.~9, pp. 1--4,
  2022.

\bibitem{o2015introduction}
K.~O'Shea and R.~Nash, ``An introduction to convolutional neural networks,''
  \emph{arXiv preprint arXiv:1511.08458}, 2015.

\bibitem{hochreiter1997long}
S.~Hochreiter and J.~Schmidhuber, ``Long short-term memory,'' \emph{Neural
  computation}, vol.~9, no.~8, pp. 1735--1780, 1997.

\bibitem{cortes1995support}
C.~Cortes and V.~Vapnik, ``Support-vector networks,'' \emph{Machine learning},
  vol.~20, no.~3, pp. 273--297, 1995.

\bibitem{breiman2001random}
L.~Breiman, ``Random forests,'' \emph{Machine learning}, vol.~45, no.~1, pp.
  5--32, 2001.

\bibitem{kingma2014adam}
D.~P. Kingma and J.~Ba, ``Adam: A method for stochastic optimization,''
  \emph{arXiv preprint arXiv:1412.6980}, 2014.

\end{thebibliography}

\end{document}